\documentclass[journal=jpcafh,manuscript=article]{achemso}
\usepackage{graphicx}
\usepackage{xcolor}
\usepackage{caption}
\usepackage[utf8]{inputenc}
\usepackage[percent]{overpic}
\usepackage[labelfont=bf]{subcaption}
\author{Dmitry V. Strelnikov}
\email{dmitry.strelnikov@kit.edu}
\affiliation{Karlsruhe Institute of Technology (KIT), Division of Physical Chemistry of Microscopic Systems, Karlsruhe, Germany}
\author{Juraj Jaš{\'i}k}
\affiliation{Department of Organic Chemistry, Faculty of Science, 
Charles University in Prague, 12843 Prague 2, Czech Republic}
\author{Dieter Gerlich}
\affiliation{Department of Physics, University of Technology, 09107 Chemnitz, Germany}
\author{Michihisa Murata}
\author{Yasujiro Murata}
\author{Koichi Komatsu}
\affiliation{Institute for Chemical Research, Kyoto University, Kyoto 611-0011, Japan}
\author{Jana Roithov{\'a}}
\affiliation{Department of Organic Chemistry, Faculty of Science, 
Charles University in Prague, 12843 Prague 2, Czech Republic}

\title{Near- and Mid-IR Gas-Phase Absorption Spectra
of H$_2$@C$_{60}^+$-He}


\begin{document}

\begin{abstract}
Near- and mid-IR absorption spectra of endohedral H$_2$@C$_{60}^+$ have been measured using He-tagging. The samples have been prepared using a  'molecular surgery' synthetic approach and were ionized and spectroscopically characterized in the gas phase. In contrast to neutral C$_{60}$ and H$_2$@C$_{60}$, the corresponding He-tagged cationic species show distinct spectral differences. Shifts and line splittings in the near- and mid-IR regions indicate the influence of the caged hydrogen molecule on both the electronic ground and excited states.  Possible relevance to astronomy is discussed.
\end{abstract}
\maketitle
\section{Introduction}
Endohedral H$_2$@C$_{60}$ can be synthesized in macroscopic quantities using a ‘molecular surgery’ approach \cite{Komatsu2005}. This has already led to numerous investigations of H$_2$@C$_{60}$, using methods such as NMR \cite{Carravetta2006,Sartori2006} or IR absorption spectroscopy \cite{Mamone2009,Room2013}. The H$_2$ inside C$_{60}$  has enough space to behave like an almost free gas-phase molecule; however, confinement leads to observable coupling of its' rotational, vibrational and translational degrees of freedom \cite{Mamone2009}.
Recently, five of the many diffuse interstellar bands (DIBs) \cite{Snow2006} have been assigned to C$_{60}^+$ using He-tagging spectroscopy in the Basel cryogenic ion trap \cite{Campbell2016}. This observation together with the fact that hydrogen is the most abundant element in the Universe, raises the questions of whether endohedral H$_2$@C$_{60}$ might be formed in Space as well and under which conditions \cite{Omont2016}. In contrast to the exohedral van der Waals complexes, endohedral H$_2$@C$_{60}^+$ can survive under ionizing conditions and higher temperatures, possibly present in some regions of Space. This stability of H$_2$@C$_{60}^+$ is demonstrated by the fact that this ion can be observed in electron ionization (EI) mass spectra of sublimed H$_2$@C$_{60}$. In order to find out whether H$_2$@C$_{60}^+$ is present in Space and provide guidance for astronomy, we investigated optical spectroscopic signatures of H$_2$@C$_{60}^+$ in the near IR range.

\section{Experimental Methods}
The experiments have been performed using the cryogenic wire quadrupole ion trap of the Prague instrument ISORI (Infrared Spectroscopy of Reaction Intermediates), described in detail in \citet{Jasik2013,Jasik2015a,Jasik2015b}. Many details concerning the production and cooling of fullerene ions and tagging them with helium atoms can be found elsewhere\cite{Gerlich2018a,Gerlich2018b}. Briefly, ions have been produced in a Finnigan Solids Probe EI-source with 65 eV electrons. Experiments have been performed with three samples, (i) 10 mg of 100\% H$_2$@C$_{60}$, (ii) 10 mg of 80\% H$_2$@C$_{60}$/20\% C$_{60}$ and (iii) C$_{60}$ (99,5\% purity). The endohedral probes have been synthesized using a ‘molecular surgery’ approach \cite{Komatsu2005}. The C$_{60}$ sample was obtained from SES Research.
Ions, emerging from the source, are mass-selected by a quadrupole mass filter and transferred via a quadrupole bender and an octopole to the cryogenic quadrupole ion trap. The temperature of the cold head has been 3.7 K. The mass-selected ions were trapped and relaxed by collisions with helium buffer gas. A part of the ions formed complexes with helium atoms. The  trapped ions were probed by IR photons from an OPO system (LaserVision)  pumped by a Nd:YAG laser (Surelite EX from Continuum, 10 Hz repetition rate, 6ns pulse width). Using an injection seeder the line width of the OPO is smaller than 1.5 cm$^{-1}$. The wavelength is measured using a high-precision wavelength meter (HighFiness WS6-200). For the near-IR electronic transitions, a Sunlite EX OPO tunable System (Continuum) is used, pumped with a seeded PL 9010 (line width $<$ 0.1 cm$^{-1}$). 
Absorption of photons was monitored by counting the number of helium complexes via a second quadrupole filter followed by an ion detector. If a complex absorbs a photon the helium is eliminated.  In order to correct for any background the light beam is blocked every other second with a mechanical shutter, resulting in numbers of unperturbed ions, $N_0$, while $N(\nu)$ is the number of complexes remaining after irradiation at frequency $\nu$. For the IR measurements $N(\nu)$ was obtained from 10 Laser pulses during one second, followed by the one second $N_0$ accumulation. For the near-IR measurements the number of the laser pulses was reduced to one. The laser pulse energy have been changing in the 0.017-0.08 mJ range for the IR measurements and in the 0.16-0.83 mJ range for the near-IR measurements. The laser beam diameters were 2 mm and 1 mm for the IR and near-IR measurements correspondingly. Laser scan velocities were 0.2 cm$^{-1}$/s for the IR and 0.03 nm/s for the near-IR.

We present the spectra either as attenuation spectra, plotting $1-N(\nu)/N_0$, or we correct in first order for non-linear attenuation by calculating relative cross section using the equation: $\sigma=-\Phi^{-1}\ln(N(\nu)/N_0)$, where $\Phi$ is number of photons per cm$^2$ the ions have been exposed to.

Raman measurements have been performed in Karlsruhe on powder samples of neutral H$_2$@C$_{60}$ and C$_{60}$ using a Raman spectrometer Kaiser Optics RXN1 (785 nm excitation, 5 cm$^{-1}$ spectral resolution). 

\section{Results and Discussion}
\subsection{Vibrational spectroscopy.}
\begin{figure*}
\includegraphics[width=\textwidth]{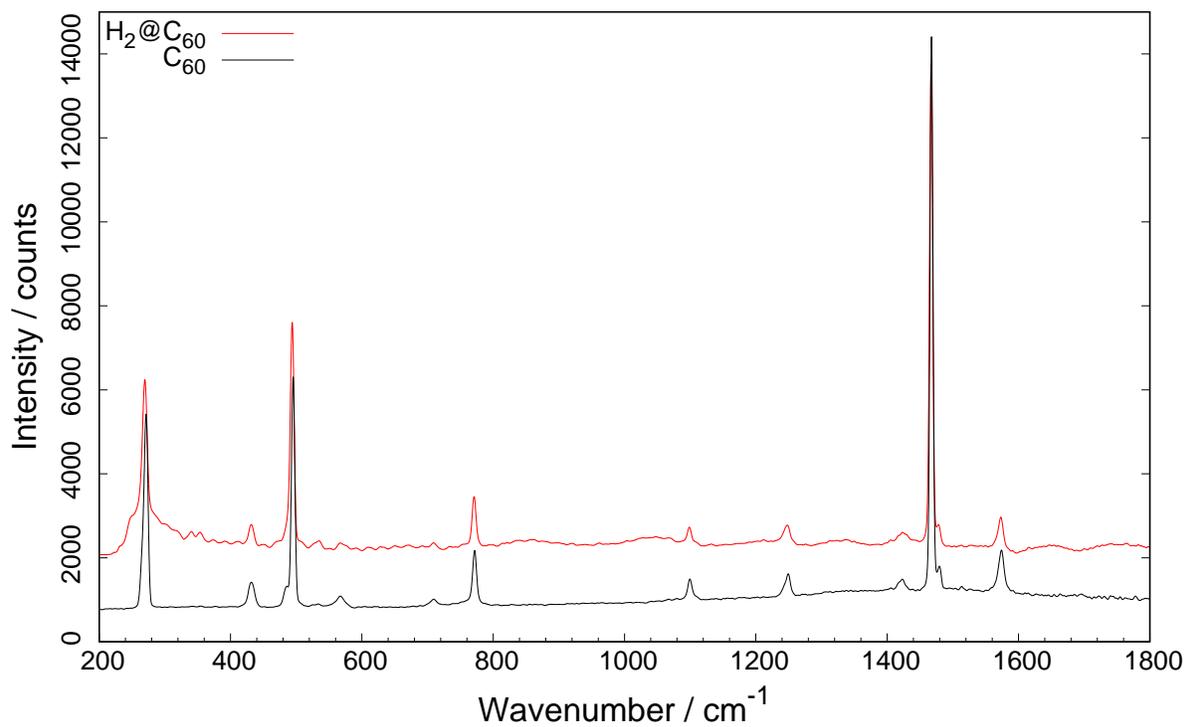}
\caption{\label{raman} Raman spectra of the bulk H$_2$@C$_{60}$ (red) and C$_{60}$ (black) samples. CW laser excitation at 785 nm, 50 mW.}
\end{figure*}

Fig. \ref{raman} compares Raman spectra of 80\% H$_2$@C$_{60}$ (red) with those from C$_{60}$ bulk samples. A careful inspection reveals that the hydrogen inside the C$_{60}$ does not influence the Raman spectrum. The line positions agree within 0.4 cm$^{-1}$. Since Raman scattering probes the vibrational levels in the electronic ground state, the state of the neutral endohedral C$_{60}$ is the same as that of the empty C$_{60}$. Some weak interference in the H$_2$@C$_{60}$ spectrum is caused by the luminescence of traces of impurities. (The interference is an instrumental artifact. It could be clearly observed when measuring continuous spectra. Probably, it is generated in the optical fiber of the Raman probe.)

\begin{figure*}
\includegraphics[width=\textwidth]{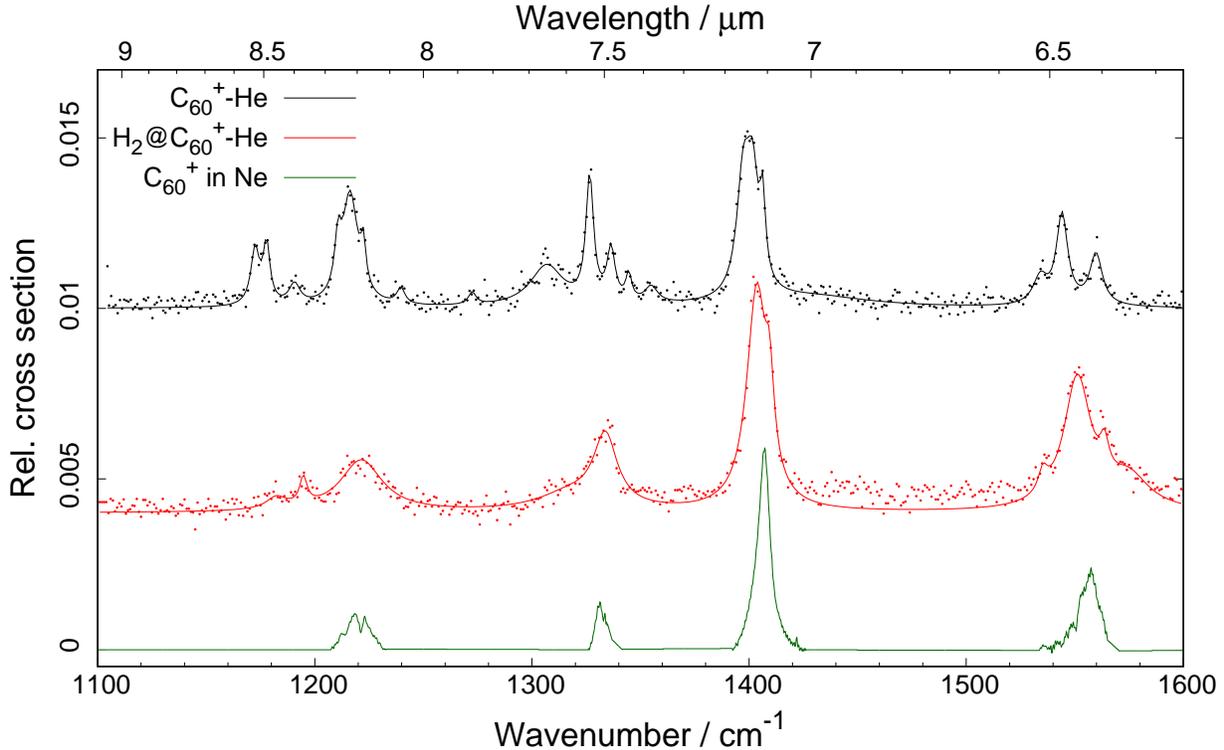}
\caption{\label{IR} IR absorption spectra of H$_2$@C$_{60}^+$-He (middle panel) compared with data  for C$_{60}^+$-He \cite{Gerlich2018a} and for C$_{60}^+$ in Ne matrix \cite{Kern2013}. The experimental data are fitted with  Lorentzians. Intensities of the H$_2$@C$_{60}^+$-He and C$_{60}^+$-He spectra are scaled down for a better comparison with C$_{60}^+$ in Ne.}
\end{figure*}

Fig. \ref{IR} compares our gas-phase spectra for H$_2$@C$_{60}^+$-He with C$_{60}^+$-He from \citet{Gerlich2018a} and with C$_{60}^+$ recorded in a cryogenic Ne-matrix \cite{Kern2013}. The two gas-phase spectra differ in the bands' shifts and splittings. These differences reflect the influence of H$_2$ on the ground state of C$_{60}^+$. Contrary to the neutral systems (see above), the cations apparently interact more strongly with the H$_2$ molecule inside. The C$_{60}^+$ cation has a $D_{5d}$ ground state symmetry \cite{Kern2013}, therefore, not all directions for translational motion of H$_2$ inside the cage are equivalent. H$_2$ can move further from the fullerene center of gravity along the $C_5$ symmetry axis. In addition, a delocalized positive charge on a fullerene cage enhances the van der Waals interaction with H$_2$.

The structure of the IR absorptions in the helium tagged fullerene ions may originate from the interaction with the attached He atom or be caused by vibrational coupling of C-C modes, the details are discussed elsewhere \cite{Gerlich2018a}.

The splittings of IR bands differ for H$_2$@C$_{60}^+$-He and C$_{60}^+$-He. The H$_2$@C$_{60}^+$-He has a reduced splitting in comparison to the empty C$_{60}^+$-He. In addition, the band at 1170 cm$^{-1}$ is either very weak or absent for H$_2$@C$_{60}^+$-He. These effects are apparently induced by the distortion from H$_2$, but the theoretical calculations do not offer a reasonable qualitative explanation of this effect. Suprisingly, the IR spectrum of H$_2$@C$_{60}^+$-He resembles more the IR spectrum of C$_{60}^+$ in neon matrix than that of C$_{60}^+$-He in the gas phase. A qualitative outcome of this measurement is the following: the more perturbation C$_{60}^+$ experiences, the larger blue shift and the smaller band splitting are observed. At present, we are not able to rationalize this by the DFT calculations (see Supporting information). The Lorentzian fit parameters of the experimental data: the band positions, relative heights and FWHMs  are compared in Table \ref{IRtable}.
\begin{table*}
 \caption{\label{IRtable}Parameters of the Lorentzian fits (see Fig.\ref{IR}): positions of the IR absorptions (cm$^{-1}$), relative heights (x1000) and FWHM (cm$^{-1}$).}
\begin{tabular}{lll|lll|lll}
\hline
\hline
\multicolumn{3}{c|}{C$_{60}^+$-He $^a$}&\multicolumn{3}{|c|}{H$_2$@C$_{60}^+$-He}&\multicolumn{3}{|c}{C$_{60}^+$ in Ne $^b$}\\
$\nu$ &rel.&FWHM&$\nu$ &rel.&FWHM&$\nu$ &rel.&FWHM\\
&height&&&height&&&height&FWHM\\
\hline
1172.5&0.6&4.8&&&&&&\\
1177.8&0.7&3.9&1182.4&0.3&11.3&&&\\
1190.7&0.2&6.1&1194.7&0.8&4.1&&&\\
1210.7&0.5&3.6&&&&1212.5&0.3&4.0\\
1216.3&1.4&8.7&&&&1218.1&1.0&4.6\\
1222.4&0.5&2.8&1221.4&1.5&23.6&1223.7&0.8&4.0\\
1239.6&0.2&4.4&&&&&&\\
1272.5&0.1&4.2&&&&&&\\
1306.8&0.5&18.4&1317.4&0.5&35.5&&&\\
1326.6&1.8&4.2&1333.8&2.1&13.3&1331.1&1.2&3.2\\
1336.4&0.7&4.5&&&&1333.6&0.5&1.2\\
1344.6&0.3&3.2&&&&1335.15&0.5&1.9\\
1354.9&0.2&7.2&&&&&&\\
1397.3&1.7&7.2&1403.5&6.4&11.0&1407.12&5.5&7.4\\
1401.6&1.6&6.8&1409.4&2.3&4.7&&&\\
1406.3&1.1&3.0&&&&&&\\
1428.9&0.1&66.3&&&&&&\\
1534.1&0.4&7.9&1535.4&0.6&5.2&&&\\
1544.1&1.4&6.4&1551.2&3.8&15.1&1557.6&2.2&9.2\\
1559.8&0.8&7.0&1563.6&0.9&5.0&&&\\
&&&1574.8&0.9&23.1&&&\\
\hline
\end{tabular}
\\ \textsuperscript{\emph{a}} Data from \citet{Gerlich2018a};
\textsuperscript{\emph{b}} Data from \citet{Kern2013}.
\end{table*} 

\subsection{Electronic spectroscopy.}
The geometry of a linear quadrupole trap together with high laser fluence in our experimental setup allows relatively fast acquisition of overview spectra: the 860 -- 970 nm region have been scanned in about one hour. Comparison of our data with \citet{Campbell2016,Campbell2018} reveals minor differences in band positions (about 0.3 \AA) but pronounced deviations in band widths and intensities (see Figs. S1,S2 in Supporting Information). These deviations originate from the non-linearities, caused by a high laser power in our measurements. The high laser power leads to the saturation of strong absorption lines, but allows to see weak absorptions. Although the attenuation for many saturated bands is about 0.8 (see below), these bands show definitively a saturation effect. The experimental geometry of the ISORI setup is such, that one has a complete overlap of the laser beam and the trapped ions. Currently, the saturation effects are not completely understood and require further dedicated experiments. Despite the deviations, the fast overview scan allows to obtain reliable absorption wavelengths. An estimate of the absorption intensities is more difficult. This can be seen from the fact that we observed more spectral bands of C$_{60}^+$-He than previously reported \cite{Campbell2016,Campbell2018}. Some of the narrow lines have been not resolved, because of the  0.25 nm step wavelength sampling. For a better resolution one would need to scan slowly with a denser sampling.
\begin{figure*}
\includegraphics[width=\textwidth]{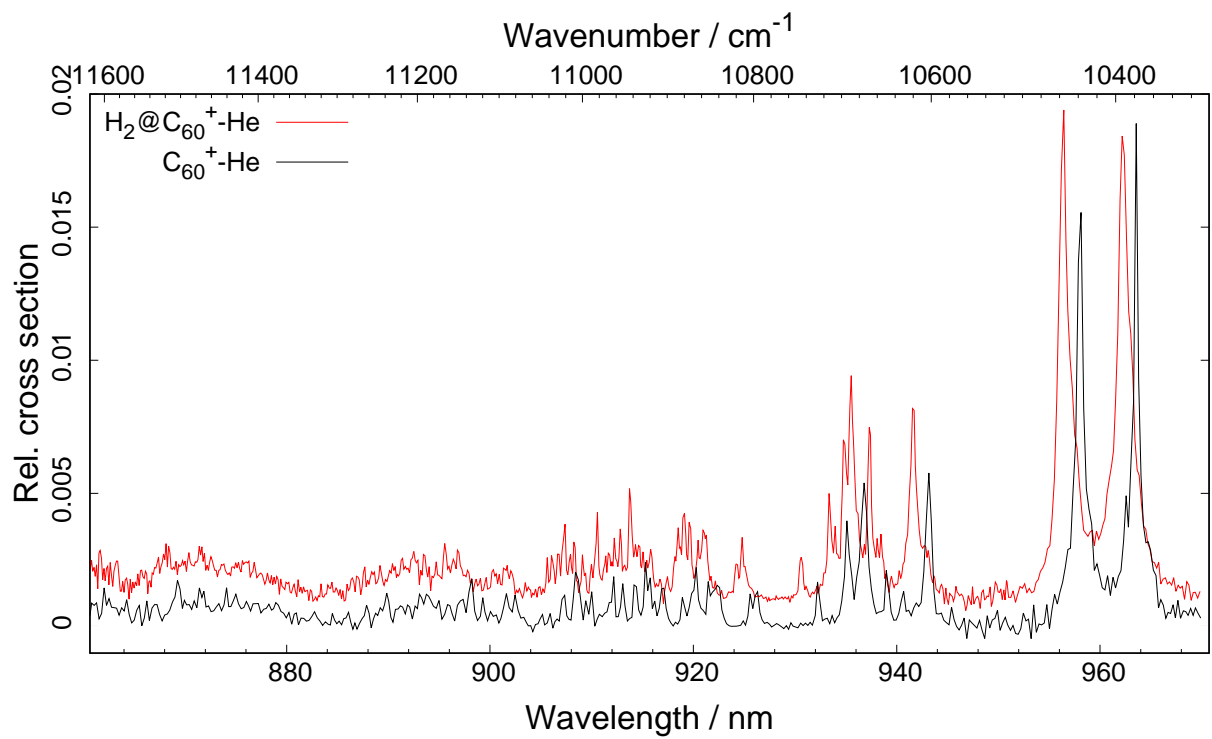}
\caption{\label{NIRsigma} An overview near-IR absorption spectra of C$_{60}^+$ vs. H$_2$@C$_{60}^+$. Both spectra are measured with the same settings. The relative line intensities of each spectrum may deviate from the unsaturated one-photon absorption spectra due to a high laser fluence.}
\end{figure*}
The electronic spectra of C$_{60}^+$-He and H$_2$@C$_{60}^+$-He show pronounced differences in band positions as well as in the band structures at higher energies (Fig. \ref{NIRsigma}). Hence, importantly the near-IR absorption spectra of C$_{60}^+$ and H$_2$@C$_{60}^+$ can be clearly distinguished. We fitted the spectra with Lorentzian functions and present the obtained fit parameters in Table \ref{NIRtable}.  Because of the limited time for the measurements, we did not obtain absolute cross sections of the observed H$_2$@C$_{60}^+$-He absorption bands, which require power dependence measurements. However, we used identical experimental settings for both measurements: C$_{60}^+$-He and H$_2$@C$_{60}^+$-He. Thus the measured absorption intensities for  both systems are comparable (Fig. \ref{NIRsigma}), and therefore the absolute absorption cross sections should be also similar.  For a direct comparison with astronomical data, one should consider a possible influence of an attached He atom on the absorption wavelengths. As in the case of C$_{60}^+$, we expect for an untagged H$_2$@C$_{60}^+$ a $\sim$0.7~\AA~blue shift\cite{Campbell2016He1-3} of the near-IR absorption wavelengths. In addition, the presented here vacuum wavelengths should be converted to the air values. Inspection of the present DIBs catalogues \cite{Hobbs2008,Hamano2016} reveals no absorption bands, corresponding to H$_2$@C$_{60}^+$. The upper limit estimation for the presence of H$_2$@C$_{60}^+$ in the interstellar medium depends on the signal to noise ratio of the astronomical data. There is a complication arising from the atmospheric water absorption lines in the near-IR region. Using the astronomical data\cite{MaierDIBs2015_2,Cox2017} we estimated the upper limit for the  H$_2$@C$_{60}^+$ to C$_{60}^+$ ratio to be $\sim$0.1.
\begin{table*}
 \caption{\label{NIRtable}Parameters of the Lorentzian fits (see Fig.3): positions of the NIR absorptions ($\lambda$ / nm), relative heights and FWHM (nm). Wavelengths are given in vacuum values.}
\begin{tabular}{lll|lll}
\hline
\hline
\multicolumn{3}{c|}{H$_2$@C$_{60}^+$-He}&\multicolumn{3}{|c}{C$_{60}^+$-He}\\
$\lambda$ &rel.&FWHM&$\lambda$ &rel.&FWHM\\
 &height&&&height&\\
\hline
962.33&0.0165&1.64&963.57&0.0170&0.68\\
956.42&0.0171&1.27&958.04&0.0145&1.01\\
942.89&0.0013&0.93&&&\\
941.61&0.0069&0.76&943.14&0.0055&0.68\\
938.37&0.0016&0.80&940.63&0.0010&0.39\\
937.33&0.0061&0.41&939.01&0.0014&0.54\\
935.55&0.0073&0.81&936.80&0.0049&0.85\\
934.80&0.0048&0.33&&&\\
933.38&0.0035&0.39&935.14&0.0032&0.63\\
930.60&0.0016&0.38&932.25&0.0016&0.27\\
924.85&0.0019&0.56&926.24&0.0018&0.23\\
921.07&0.0025&0.79&922.43&0.0015&0.45\\
919.63&0.0023&0.29&921.48&0.0013&0.31\\
919.08&0.0031&0.41&920.29&0.0019&0.40\\
918.49&0.0022&0.34&919.79&0.0007&0.49\\
915.83&0.0017&0.43&916.98&0.0014&0.52\\
913.77&0.0041&0.34&915.31&0.0020&0.31\\
912.83&0.0022&0.34&914.29&0.0016&0.40\\
910.53&0.0031&0.27&912.16&0.0015&0.28\\
907.31&0.0028&0.33&908.50&0.0020&0.59\\
\hline
\end{tabular} 
\end{table*} 

\begin{figure*}
\includegraphics[width=\textwidth]{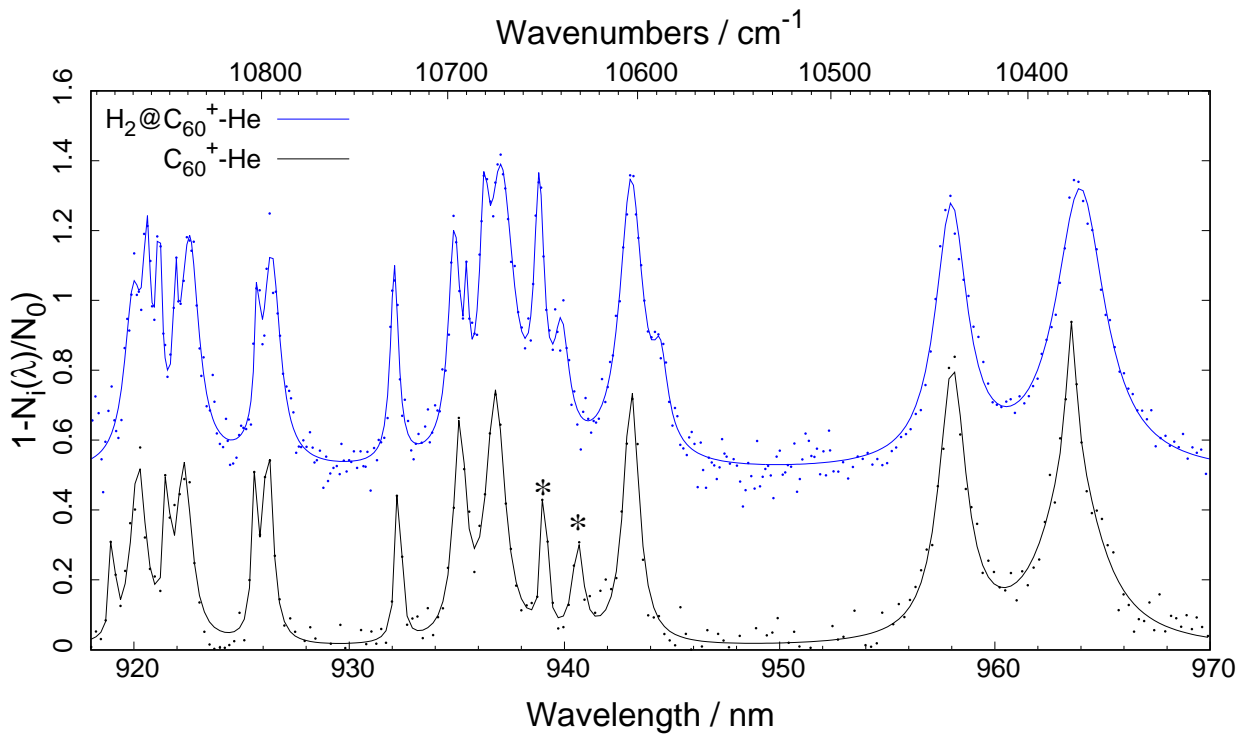}
\caption{\label{NIRatt} Near-IR photo-depletion spectra of C$_{60}^+$ vs. H$_2$@C$_{60}^+$. N($\lambda$) is the number of survived He-complexes after the laser irradiation, N$_0$ is the number of the He-complexes in the absence of laser. The  H$_2$@C$_{60}^+$ spectrum is shifted by 1.5 nm to the longer wavelengths for an easier comparison of both spectra. ‘*’ indicate newly observed weak absorption bands of C$_{60}^+$.}
\end{figure*}
Fig.\ref{NIRatt} shows a section from Fig. \ref{NIRsigma} as attenuation. For better comparison, the H$_2$@C$_{60}^+$-He spectrum has been blue-shifted by about 1.5 nm with respect to the C$_{60}^+$-He spectrum. Detailed inspection reveals that some of the lines in the H$_2$@C$_{60}^+$-He spectrum exhibit  some splitting. Apparently, H$_2$ perturbs the excited state of C$_{60}^+$. This is, however, not surprising, since the ground state is also influenced by the presence of H$_2$ inside the cage. So far there is no theoretical work, properly describing and explaining the vibronic transitions of C$_{60}^+$ in the $^2$E$_{1g}$ state. With a hydrogen molecule inside the cage it is an even more challenging system for a theoretical treatment than C$_{60}^+$. TDDFT was done for several different orientations of H$_2$. The theory predicts the blue shifts up to 2.5 nm and also the splitting for some of the H$_2$ positions (see Supporting Information). 
Line shifts in the electronic spectrum, introduced by H$_2$ inside a fullerene are much larger, than those, induced by adsorption of Ne, Ar, N$_2$, H$_2$, D$_2$ outside the C$_{60}^+$ \cite{Holz2017}. The external molecule/atom adsorption on C$_{60}^+$ molecules mostly leads to red shifts in the near-IR spectra \cite{Holz2017}. Conversely, we observe a considerable blue shift in the spectrum of the endohedral H$_2$@C$_{60}^+$ ion. 
\section{Conclusion}
Gas-phase mid-IR and near-IR measurements show distinct spectral differences between H$_2$@C$_{60}^+$ and C$_{60}^+$. The correct quantitative description of the observed differences in the absorptions requires dedicated theoretical modeling. There are no known DIBs at the absorption positions of H$_2$@C$_{60}^+$. The estimated upper limit of the interstellar H$_2$@C$_{60}^+$/C$_{60}^+$ ratio is $\sim$0.1.

\begin{acknowledgement}
The project was funded by the European Research Council (ERC CoG No. 682275) and the Deutsche Forschungsgemeinschaft (KA 972/10-1). We also acknowledge support by KIT and Land Baden-W\"{u}rttemberg. 
\end{acknowledgement}

%\bibliography{H2atC60}

\begin{thebibliography}{1}
\bibitem[Komatsu \latin{et~al.}(2005)Komatsu, Murata, and Murata]{Komatsu2005}
Komatsu,~K.; Murata,~M.; Murata,~Y. Encapsulation of Molecular Hydrogen in
  Fullerene C60 by Organic Synthesis. \emph{Science} \textbf{2005}, \emph{307},
  238--240
\bibitem[Carravetta \latin{et~al.}(2006)Carravetta, Johannessen, Levitt,
  Heinmaa, Stern, Samoson, Horsewill, Murata, and Komatsu]{Carravetta2006}
Carravetta,~M.; Johannessen,~O.~G.; Levitt,~M.~H.; Heinmaa,~I.; Stern,~R.;
  Samoson,~A.; Horsewill,~A.~J.; Murata,~Y.; Komatsu,~K. Cryogenic NMR
  Spectroscopy of Endohedral Hydrogen-Fullerene Complexes. \emph{J. Chem.
  Phys.} \textbf{2006}, \emph{124}, 104507
\bibitem[Sartori \latin{et~al.}(2006)Sartori, Ruzzi, Turro, Decatur,
  Doetschman, Lawler, Buchachenko, Murata, and Komatsu]{Sartori2006}
Sartori,~E.; Ruzzi,~M.; Turro,~N.~J.; Decatur,~J.~D.; Doetschman,~D.~C.;
  Lawler,~R.~G.; Buchachenko,~A.~L.; Murata,~Y.; Komatsu,~K. Nuclear Relaxation
  of H2 and H2@C60 in Organic Solvents. \emph{J. Am. Chem. Soc.} \textbf{2006},
  \emph{128}, 14752--14753, PMID: 17105254
\bibitem[Mamone \latin{et~al.}(2009)Mamone, Ge, Hüvonen, Nagel, Danquigny,
  Cuda, Grossel, Murata, Komatsu, Levitt, R{\~o}{\~o}m, and
  Carravetta]{Mamone2009}
Mamone,~S.; Ge,~M.; Hüvonen,~D.; Nagel,~U.; Danquigny,~A.; Cuda,~F.;
  Grossel,~M.~C.; Murata,~Y.; Komatsu,~K.; Levitt,~M.~H. \latin{et~al.}  Rotor
  in a Cage: Infrared Spectroscopy of an Endohedral Hydrogen-Fullerene Complex.
  \emph{J. Chem. Phys.} \textbf{2009}, \emph{130}, 081103
\bibitem[R{\~o}{\~o}m \latin{et~al.}(2013)R{\~o}{\~o}m, Peedu, Ge, H{\"u}vonen,
  Nagel, Ye, Xu, Ba{\v c}i{\'c}, Mamone, Levitt, Carravetta, Chen, Lei, Turro,
  Murata, and Komatsu]{Room2013}
R{\~o}{\~o}m,~T.; Peedu,~L.; Ge,~M.; H{\"u}vonen,~D.; Nagel,~U.; Ye,~S.;
  Xu,~M.; Ba{\v c}i{\'c},~Z.; Mamone,~S.; Levitt,~M.~H. \latin{et~al.}
  Infrared Spectroscopy of Small-Molecule Endofullerenes. \emph{Philos. Trans.
  R. Soc., A} \textbf{2013}, \emph{371}
\bibitem[Snow and McCall(2006)Snow, and McCall]{Snow2006}
Snow,~T.~P.; McCall,~B.~J. Diffuse Atomic and Molecular Clouds. \emph{Annu.
  Rev. Astron. Astrophys.} \textbf{2006}, \emph{44}, 367--414
\bibitem[Campbell \latin{et~al.}(2016)Campbell, Holz, Maier, Gerlich, Walker,
  and Bohlender]{Campbell2016}
Campbell,~E.~K.; Holz,~M.; Maier,~J.~P.; Gerlich,~D.; Walker,~G. A.~H.;
  Bohlender,~D. Gas Phase Absorption Spectroscopy of {C}$_{60}^+$
  and{C}$_{70}^+$ in a Cryogenic Ion Trap: Comparison with Astronomical
  Measurements. \emph{Astrophys. J.} \textbf{2016}, \emph{822}, 17
\bibitem[Omont(2016)]{Omont2016}
Omont,~A. Interstellar fullerene compounds and diffuse interstellar bands.
  \emph{Astron. Astrophys.} \textbf{2016}, \emph{590}, A52
\bibitem[Jaš\'{i}k \latin{et~al.}(2013)Jaš\'{i}k, Žabka, Roithov\'{a}, and
  Gerlich]{Jasik2013}
Jaš\'{i}k,~J.; Žabka,~J.; Roithov\'{a},~J.; Gerlich,~D. Infrared Spectroscopy
  Of Trapped Molecular Dications Below 4K. \emph{Int. J. Mass Spectrom.}
  \textbf{2013}, \emph{354-355}, 204 -- 210, Detlef Schröder Memorial
  Issue
\bibitem[Jaš\'{i}k \latin{et~al.}(2015)Jaš\'{i}k, Gerlich, and
  Roithov\'{a}]{Jasik2015a}
Jaš\'{i}k,~J.; Gerlich,~D.; Roithov\'{a},~J. Two-Color Infrared
  Predissociation Spectroscopy of C$_6$H$_6^{2+}$ Isomers Using Helium Tagging.
  \emph{J. Phys. Chem. A} \textbf{2015}, \emph{119}, 2532--2542, PMID:
  25402726
\bibitem[Jaš\'{i}k \latin{et~al.}(2015)Jaš\'{i}k, Navrátil, Němec, and
  Roithov\'{a}]{Jasik2015b}
Jaš\'{i}k,~J.; Navrátil,~R.; Němec,~I.; Roithov\'{a},~J. Infrared and
  Visible Photodissociation Spectra of Rhodamine Ions at 3 K in the Gas Phase.
  \emph{J. Phys. Chem. A} \textbf{2015}, \emph{119}, 12648--12655, PMID:
  26595323
\bibitem[Gerlich \latin{et~al.}(2018)Gerlich, Jaš\'{i}k, Strelnikov, and
  Roithov\'{a}]{Gerlich2018a}
Gerlich,~D.; Jaš\'{i}k,~J.; Strelnikov,~D.; Roithov\'{a},~J. IR spectroscopy
  of fullerene ions in a cryogenic quadrupole trap. \textbf{2018}, Astrophys.
  J., in press
\bibitem[Gerlich \latin{et~al.}(2018)Gerlich, Jaš{\'i}k, and
  Roithov{\'a}]{Gerlich2018b}
Gerlich,~D.; Jaš{\'i}k,~J.; Roithov{\'a},~J. Tagging fullerene ions with
  helium in a cryogenic quadrupole trap. \textbf{2018}, In preparation for the
  special issue of the International Journal of Mass Spectrometry to honor
  Professor Helmut Schwarz
\bibitem[Kern \latin{et~al.}(2013)Kern, Strelnikov, Weis, B\"{o}ttcher, and
  Kappes]{Kern2013}
Kern,~B.; Strelnikov,~D.; Weis,~P.; B\"{o}ttcher,~A.; Kappes,~M.~M. {IR}
  Absorptions of {C}$_{60}^+$ and {C}$_{60}^-$ in Neon Matrixes. \emph{J. Phys.
  Chem. A} \textbf{2013}, \emph{117}, 8251--8255
\bibitem[Campbell and Maier(2018)Campbell, and Maier]{Campbell2018}
Campbell,~E.~K.; Maier,~J.~P. Isomeric and Isotopic Effects on the Electronic
  Spectrum of C60+-He: Consequences for Astronomical Observations of C60+.
  \emph{Astrophys. J.} \textbf{2018}, \emph{858}, 36
\bibitem[Campbell \latin{et~al.}(2016)Campbell, Holz, and
  Maier]{Campbell2016He1-3}
Campbell,~E.~K.; Holz,~M.; Maier,~J.~P. C60+ in Diffuse Clouds: Laboratory and
  Astronomical Comparison. \emph{Astrophys. J., Lett.} \textbf{2016},
  \emph{826}, L4
\bibitem[{Hobbs} \latin{et~al.}(2008){Hobbs}, {York}, {Snow}, {Oka},
  {Thorburn}, {Bishof}, {Friedman}, {McCall}, {Rachford}, {Sonnentrucker}, and
  {Welty}]{Hobbs2008}
{Hobbs},~L.~M.; {York},~D.~G.; {Snow},~T.~P.; {Oka},~T.; {Thorburn},~J.~A.;
  {Bishof},~M.; {Friedman},~S.~D.; {McCall},~B.~J.; {Rachford},~B.;
  {Sonnentrucker},~P. \latin{et~al.}  {A Catalog of Diffuse Interstellar Bands
  in the Spectrum of HD 204827}. \emph{Astrophys. J.} \textbf{2008},
  \emph{680}, 1256--1270
\bibitem[{Hamano} \latin{et~al.}(2016){Hamano}, {Kobayashi}, {Kondo},
  {Sameshima}, {Nakanishi}, {Ikeda}, {Yasui}, {Mizumoto}, {Matsunaga}, {Fukue},
  {Yamamoto}, {Izumi}, {Mito}, {Nakaoka}, {Kawanishi}, {Kitano}, {Otsubo},
  {Kinoshita}, and {Kawakita}]{Hamano2016}
{Hamano},~S.; {Kobayashi},~N.; {Kondo},~S.; {Sameshima},~H.; {Nakanishi},~K.;
  {Ikeda},~Y.; {Yasui},~C.; {Mizumoto},~M.; {Matsunaga},~N.; {Fukue},~K.
  \latin{et~al.}  {Near Infrared Diffuse Interstellar Bands Toward the Cygnus
  OB2 Association}. \emph{Astrophys. J.} \textbf{2016}, \emph{821}, 42
\bibitem[Walker \latin{et~al.}(2015)Walker, Bohlender, Maier, and
  Campbell]{MaierDIBs2015_2}
Walker,~G. A.~H.; Bohlender,~D.~A.; Maier,~J.~P.; Campbell,~E.~K.
  Identification of More Interstellar {C}$_{60}^+$ Bands. \emph{Astrophys. J.,
  Lett.} \textbf{2015}, \emph{812}, L8
\bibitem[{Cox, Nick L. J.} \latin{et~al.}(2017){Cox, Nick L. J.}, {Cami, Jan},
  {Farhang, Amin}, {Smoker, Jonathan}, {Monreal-Ibero, Ana}, {Lallement,
  Rosine}, {Sarre, Peter J.}, {Marshall, Charlotte C. M.}, {Smith, Keith T.},
  {Evans, Christopher J.}, {Royer, Pierre}, {Linnartz, Harold}, {Cordiner,
  Martin A.}, {Joblin, Christine}, {van Loon, Jacco Th.}, {Foing, Bernard H.},
  {Bhatt, Neil H.}, {Bron, Emeric}, {Elyajouri, Meriem}, {de Koter, Alex},
  {Ehrenfreund, Pascale}, {Javadi, Atefeh}, {Kaper, Lex}, {Khosroshadi, Habib
  G.}, {Laverick, Mike}, {Le Petit, Franck}, {Mulas, Giacomo}, {Roueff,
  Evelyne}, {Salama, Farid}, and {Spaans, Marco}]{Cox2017}
{Cox, Nick L. J.},; {Cami, Jan},; {Farhang, Amin},; {Smoker, Jonathan},;
  {Monreal-Ibero, Ana},; {Lallement, Rosine},; {Sarre, Peter J.},; {Marshall,
  Charlotte C. M.},; {Smith, Keith T.},; {Evans, Christopher J.},
  \latin{et~al.}  The ESO Diffuse Interstellar Bands Large Exploration Survey
  (EDIBLES) - I. Project description, survey sample, and quality assessment.
  \emph{Astron. Astrophys.} \textbf{2017}, \emph{606}, A76
\bibitem[Holz \latin{et~al.}(2017)Holz, Campbell, Rice, and Maier]{Holz2017}
Holz,~M.; Campbell,~E.~K.; Rice,~C.~A.; Maier,~J.~P. Electronic Absorption
  Spectra of C60+–L (L=He, Ne, Ar, Kr, H2, D2, N2) Complexes. \emph{J. Mol.
  Spectrosc.} \textbf{2017}, \emph{332}, 22 -- 25, Molecular Spectroscopy in
  Traps
\bibitem[Rohatgi()]{WebPlotDigitizer}
Rohatgi,~A. WebPlotDigitizer. \url{https://automeris.io/WebPlotDigitizer},
  Version: 4.1, E-Mail: ankitrohatgi@hotmail.com, Location: Austin, Texas,
  USA
\bibitem[Tur()]{Turbomole}
{TURBOMOLE V6.4 2012, a development of University of Karlsruhe and
  Forschungszentrum Karlsruhe GmbH, 1989-2007, TURBOMOLE GmbH, since 2007;
  available from http://www.turbomole.com}
\end{thebibliography}

\clearpage

\section*{Supporting Information} 
\begin{figure*}
\includegraphics[width=\textwidth]{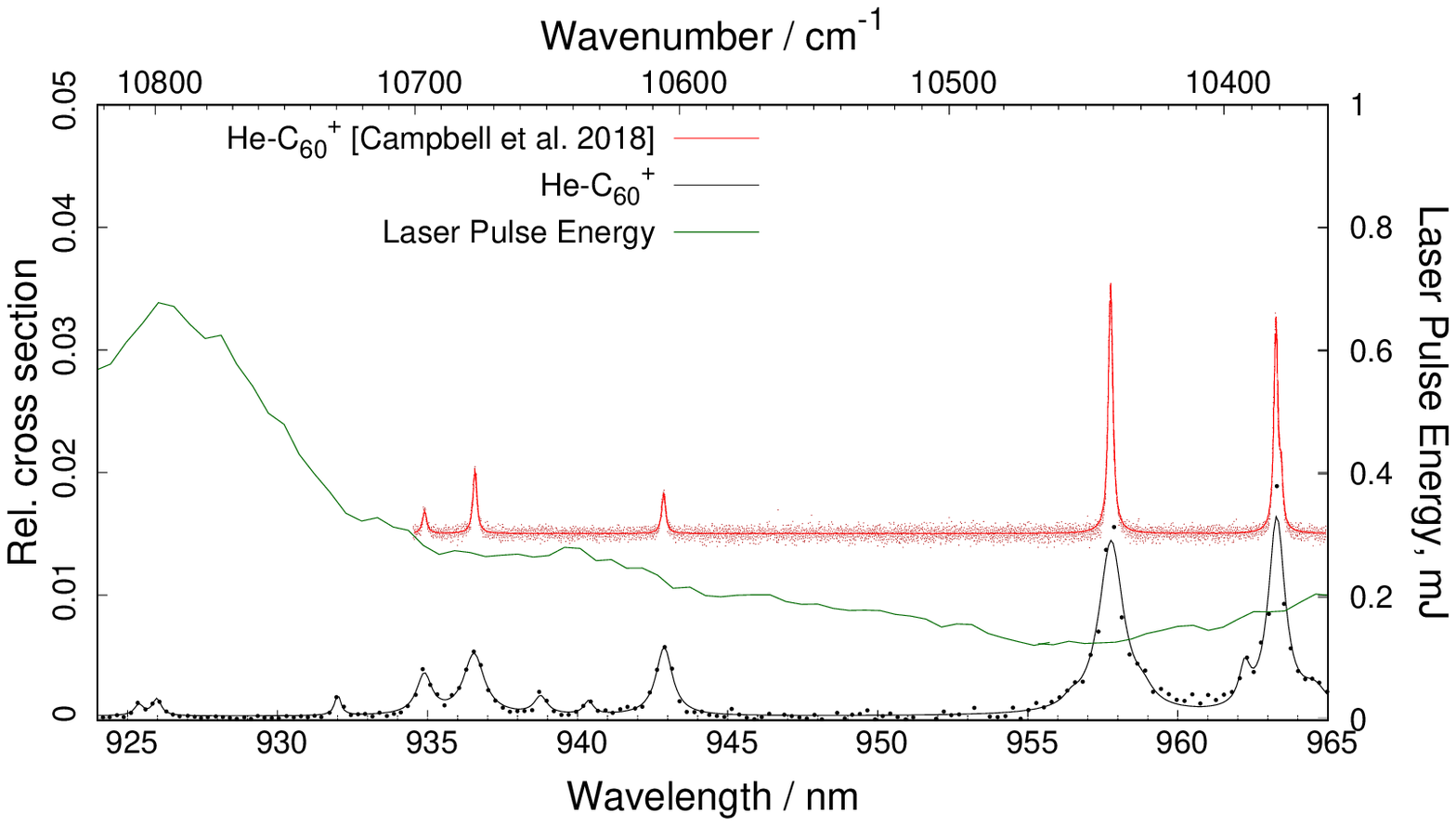}
\caption*{Figure S1: Comparison of our near-IR data to the \citet{Campbell2018}. Here the vacuum wavelengths were converted to the air wavelengths, used in \citet{Campbell2018}. The data points from \citet{Campbell2018} have been digitized using \textit{WebPlotDigitizer}\cite{WebPlotDigitizer}.}\label{nircampb1}
\end{figure*}
\begin{figure*}
\includegraphics[width=\textwidth]{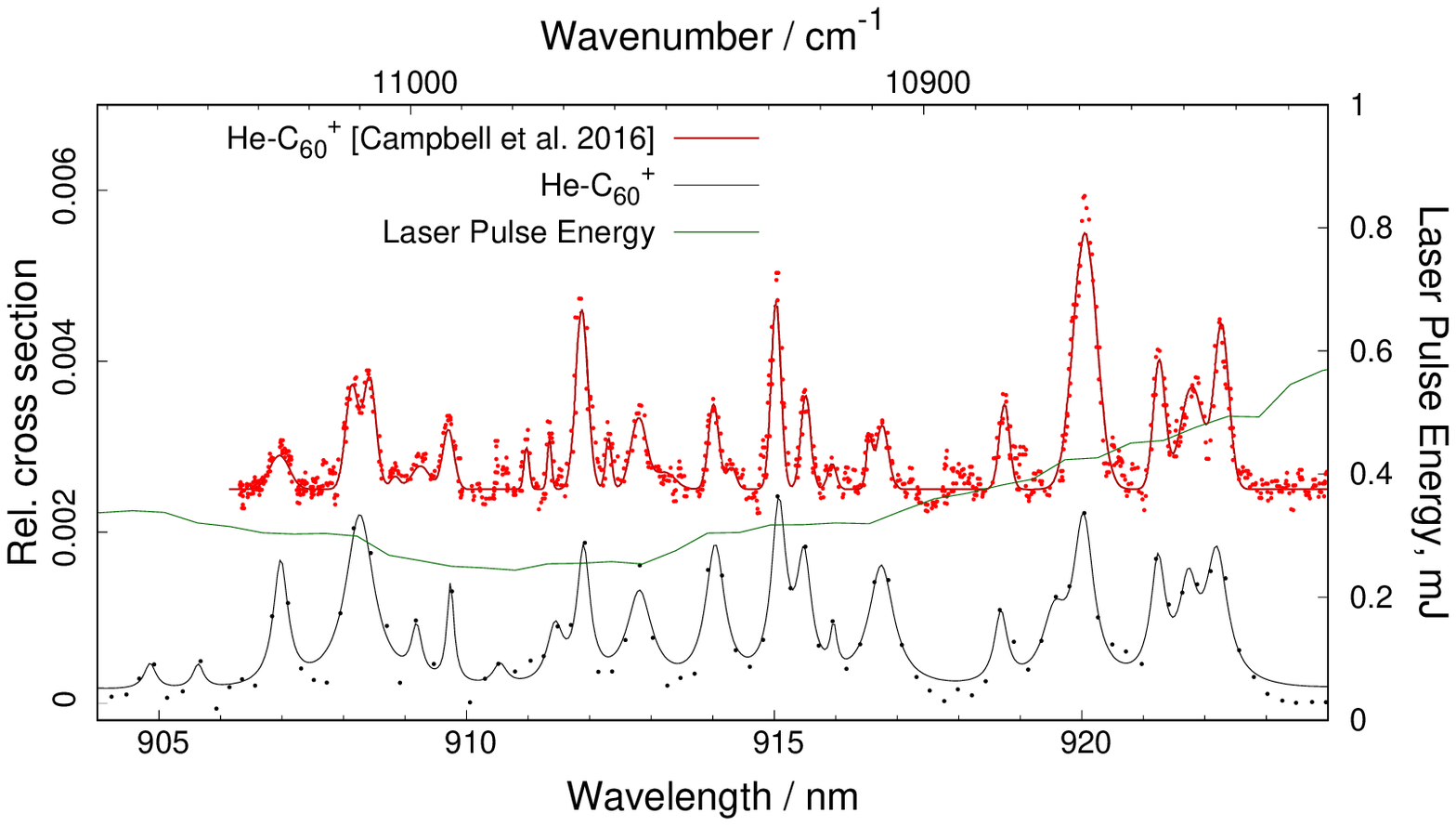}
\caption*{Figure S2: Comparison of our near-IR data to the \citet{Campbell2016}. Here the vacuum wavelengths were converted to the air wavelengths used in \citet{Campbell2016}.The data points from \citet{Campbell2016} have been digitized using \textit{WebPlotDigitizer}\cite{WebPlotDigitizer}.}\label{nircampb}
\end{figure*}
\begin{figure*}
\includegraphics[width=\textwidth]{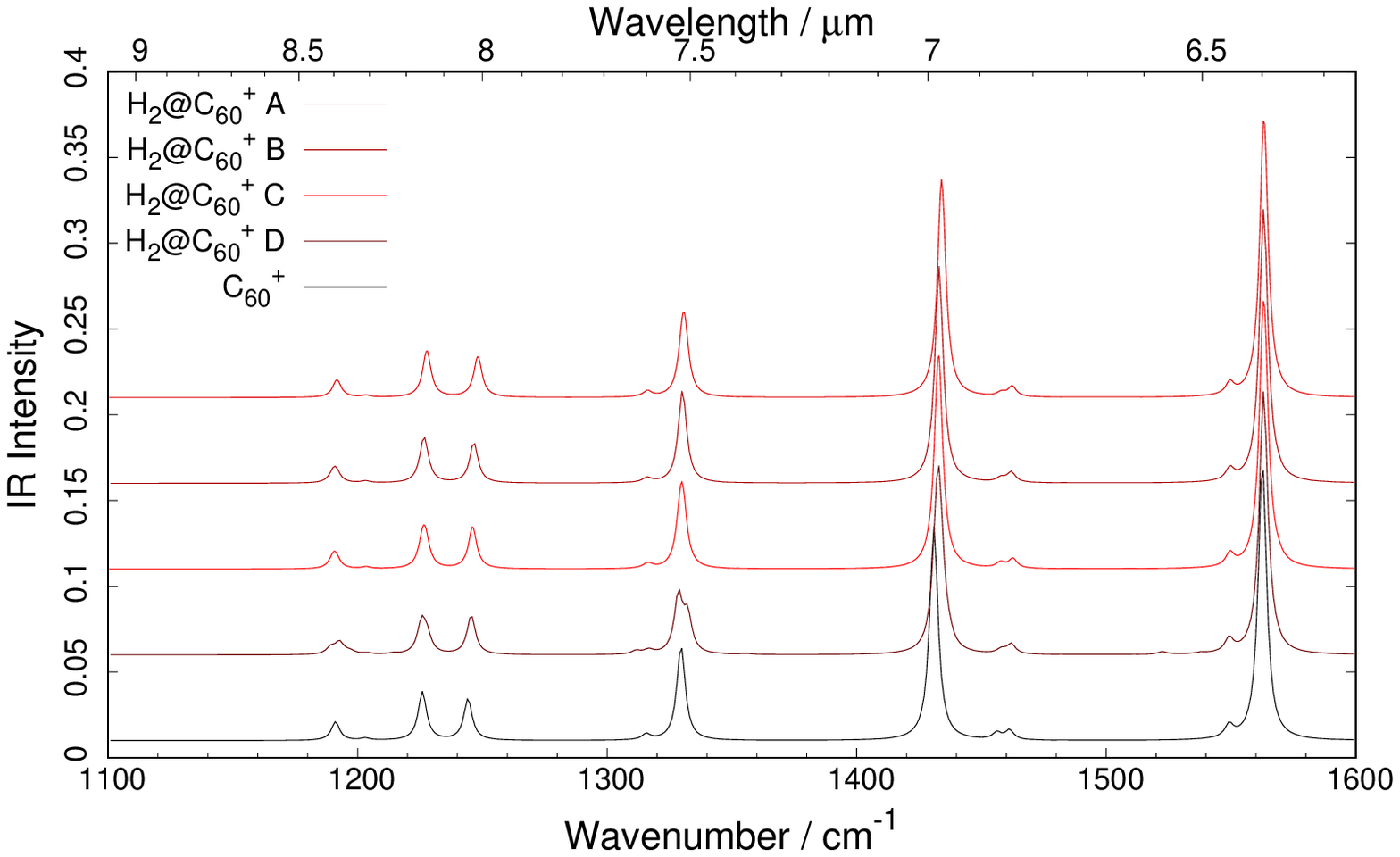}
\caption*{Figure S3: Calculated harmonic frequencies (RI-DFT BP86/def2-SVP, D3 dispersion correction \cite{Turbomole}) of isomers with different H$_2$ orientations inside C$_{60}^+$. The calculated spectrum was broadened by Lorentzian functions with a 4 cm$^{-1}$ FWHM. The potential energy surface for H$_2$ inside C$_{60}^+$ is very shallow, hindering a proper geometry optimization. Therefore, we took few H$_2$@C$_{60}^+$ isomers, which are not the real ground states due to the presence of imaginary frequencies, corresponding to H$_2$ motion inside C$_{60}^+$. Nevertheless, the applied level of theory does not predict considerable influence of H$_2$ on the C-C stretching (tangential) vibrational modes.}\label{dftir}
\end{figure*}
\begin{figure*}
\includegraphics[width=\textwidth]{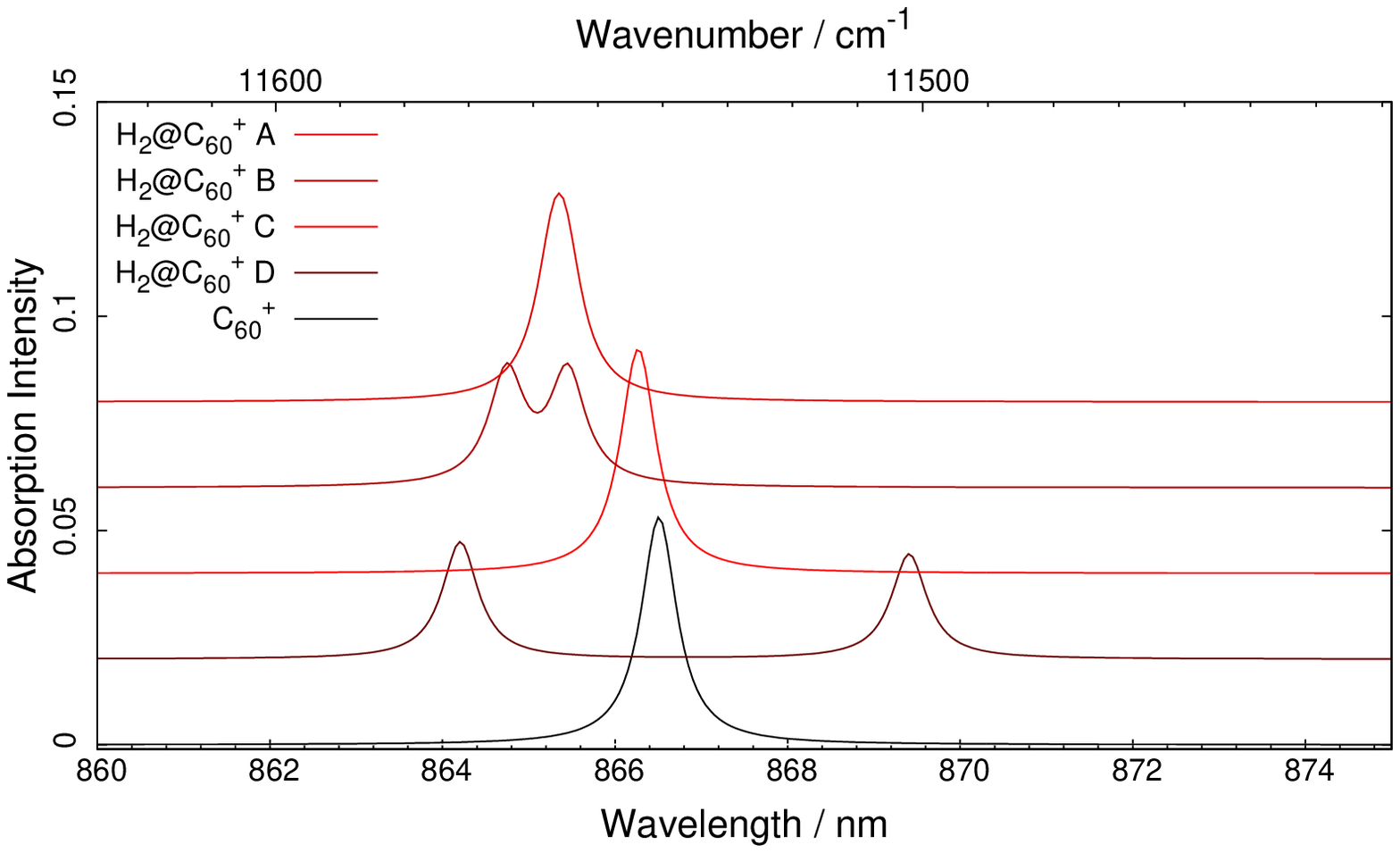}
\caption*{Figure S4: Calculated 0-0 vertical electronic transitions, corresponding to the C$_{60}^+$ ${^2E_{1g}\leftarrow {^2A_{1u}}}$ absorption (TDDFT BP86/def2-SVP)\cite{Turbomole} for the same H$_2$@C$_{60}^+$ isomers as in Fig. S3. The calculated spectrum was broadened by Lorentzian functions with a 0.5 nm FWHM: oscillator strength x Lorentzian function. Most of the calculated absorption wavelengths show a blue shift relative to the C$_{60}^+$ absorption, which is also observed in the experiment.}\label{tddft}
\end{figure*}

\end{document}